\documentclass[%
 iop,
 jpd,%
 amsmath,amssymb,
reprint,%
floatfix]{revtex4-1}

\usepackage{graphicx}
\usepackage{dcolumn}
\usepackage{bm}
\usepackage[mathlines]{lineno}

\begin{document}

\title[\title{Magnetic nanoparticle sensing}
]{Magnetic nanoparticle sensing: decoupling the magnetization from the excitation field}

\author{Daniel B. Reeves}
 \affiliation{Department of Physics and Astronomy, Dartmouth College, 6127 Wilder Hall, Hanover, NH 03755}
 \email{dbr@Dartmouth.edu.}
 
\author{John B. Weaver}%
\altaffiliation[Also at ]{Radiology Department, Geisel School of Medicine, Hanover, NH 03755}

\date{Jan. 2014}

\begin{abstract} Remote sensing of magnetic nanoparticles has exciting applications for magnetic nanoparticle hyperthermia and molecular detection. We introduce, simulate, and experimentally demonstrate an innovation---a sensing coil that is geometrically decoupled from the excitation field---for magnetic nanoparticle spectroscopy that increases the flexibility and capabilities of remote detection. The decoupling enhances the sensitivity absolutely; to small amounts of nanoparticles, and relatively; to small changes in the nanoparticle dynamics. We adapt a previous spectroscopic method that measures the relaxation time of nanoparticles and demonstrate a new measurement of nanoparticle temperature that could potentially be used concurrently during hyperthermia.  \end{abstract}

\maketitle

\section{Motivation}

\subsection{Sensing nanoparticles}

Magnetic nanoparticles (MNPs) are currently used as sensors in a wide range of applications \cite{KOH,HAUN}. They are particularly useful because their concentration and properties can be measured remotely through the strength of their magnetization \cite{SEIKO}. Coupling this with advances in specific chemical binding, MNPs can also be used to measure analyte concentration \cite{XJ} or to enable magnetic separation \cite{PANK}. Molecular concentration sensing with MNPs has the exciting potential to provide a non-invasive \emph{in vivo} assay that could track the concentration of a selected molecule over time. MNP molecular sensing is very sensitive (100pM analyte concentrations can be detected) and has distinct abilities for \emph{in vivo} sensing \cite{XJ}.  Additionally, MNPs have been used to sense other parameters including temperature \cite{TEMPEST}, viscosity \cite{VISC}, and rigidity of a cellular matrix \cite{RIGID}.

Detection of magnetization harmonics from MNPs excited by an applied sinusoidal magnetic field is referred to as magnetic nanoparticle spectroscopy (MPS). In the regime of physical Brownian rotation, nanoparticle rotations connect the MNP dynamics to the local microenvironment. The average rotational freedom of the MNPs can be used as a surrogate measurement for environmental parameters; for instance, in molecular concentration sensing, the rotational freedom indicates the number of MNPs that are bound by a selected analyte. In the N\'{e}el regime, where particles are fixed in space with magnetic moments that switch internally, the spectra carry information about solid-state parameters like anisotropy or temperature.

\subsection{Monitoring temperature during hyperthermia therapy}

Magnetic fluid hyperthermia (MFH) is already well developed and considered a promising addition to current cancer therapies \cite{HERGT,JORDAN,HYPERTHERM}. The technology employs remote rotation of magnetic nanoparticles (MNPs) to deliver specifically directed cytotoxic heat. Significant therapeutic effects have been demonstrated, yet general application remains out of reach because it is virtually impossible to predict the temperature that will be achieved \cite{BIHAN,TCC} and adequate heating often requires unsafe clinical practices \cite{PANK,MONITOR}. Several other issues preventing clinical implementation have been known for years, and are outlined well by Cetas: 1) interference of measurement with therapy; 2) interference of excitation field with measurements; 3) resolution; and 4) challenges of decoupling temperature changes from physiological effects \cite{TCC}. 

Modeling of simple systems is essential to reach a level of understanding whereby heat deposition during MFH is predictable. Still, current models are insufficient because \emph{in vivo} issues (e.g., heat diffusion through blood flow, immune response, cellular binding) complicate the physics to a degree far beyond current theoretical abilities. Moreover, experiments \emph{in vitro} are simply inadequate to predict MNP heating in complicated physiology.

Noninvasive thermometry has been accomplished using magnetic resonance imaging \cite{BIHAN}, but field interactions prevent concurrent use during MFH and therefore the first two of the listed problems remain. An alternative to conventional imaging requires a shift of perspective. Instead of measuring the local tissue temperature around the MNPs, it is possible to infer the temperature of the MNPs \emph{themselves} through remote energy measurements \cite{TEMPEST}. MPS is an attractive candidate for this purpose whereby the same MNPs used to deliver heat in hyperthermia can be simultaneously monitored with induction coils. In this paper, we describe a new spectroscopic method capable of providing information about nanoparticle magnetizations, temperatures, and rotational freedom with high sensitivity due to a theoretically complete decoupling of the sensing coil from the excitation field. We demonstrate dynamical simulations that agree with prior MPS experimental results, and introduce a new method to measure the temperature of magnetic nanoparticles that exploits the new spectroscopic method.

\section{Magnetic nanoparticle spectroscopy (MPS) in theory}

Magnetic nanoparticle spectroscopy (MPS) has been previously discussed as a method to gather remote information about nanoparticles through their harmonic spectra. By using MNPs in the upper range of the nanoscale (e.g., 100nm) we can guarantee MNP coupling to the microenvironment while still retaining nanomagnetic properties. The larger particles should be thermally blocked\cite{WFB} and indeed in previous studies we find that changing parameters of the suspension fluid changes the relaxation time of the particles \cite{VISC}. This demonstrates the relaxation mechanism is Brownian, and thus we can use the Brownian relaxation time\cite{EIN} $\tau$, defined in terms of suspension viscosity $\eta$, particle hydrodynamic volume $V_\mathrm{hy}$,  Boltzmann's constant $k_\mathrm{B}$ and temperature $T$ to make statements about environmental parameters:
\begin{equation} \tau = \frac{3\eta V_{\mathrm{hy}}}{k_\mathrm{B}T}. \label{tau} \end{equation}
In MPS, a sinusoidally varying magnetic field is used to excite samples of MNPs. The particles subsequently radiate and induction coils are used to measure the non-linear magnetization response of the particles. The induced voltage $\mathcal{V}$ in the `pickup' or `sensing' coil is the rate-change of magnetic flux coming from oscillating MNPs
\begin{equation} \mathcal{V}(t) \propto \frac{\partial M}{\partial t}. \end{equation}
The average normalized magnetization $M$ can be expressed as a Fourier series in terms of the  fundamental drive frequency ($\omega_o=2\pi f_o$) and integer $h$ multiples of this frequency:
\begin{equation} \frac{M(t)}{|M|} =\sum_h a_h e^{ih \omega_o t}. \end{equation}
The Fourier coefficients at integer multiples of the fundamental frequency (the harmonics) are in principle independent of the excitation field, and are thus used to selectively measure the MNP magnetization. This signal $\mathcal{S}$ is the list of Fourier series components---the harmonics of the \emph{time derivative} of the magnetization
\begin{equation} \mathcal{S}(h) \propto  h|a_h| . \label{sig}\end{equation}
 Note, taking the derivative introduces the additional factor of harmonic number. The magnetization magnitude is linear with respect to MNP concentration in a dilute sample and thus a ratio of harmonics (e.g., $r_{5/3} = a_5/a_3$) is a convenient concentration-independent metric used to discuss particle dynamics. 
 
\section{Static field magnetic spectroscopy}

\begin{figure}[h!] \begin{center} \includegraphics[width=3.25in]{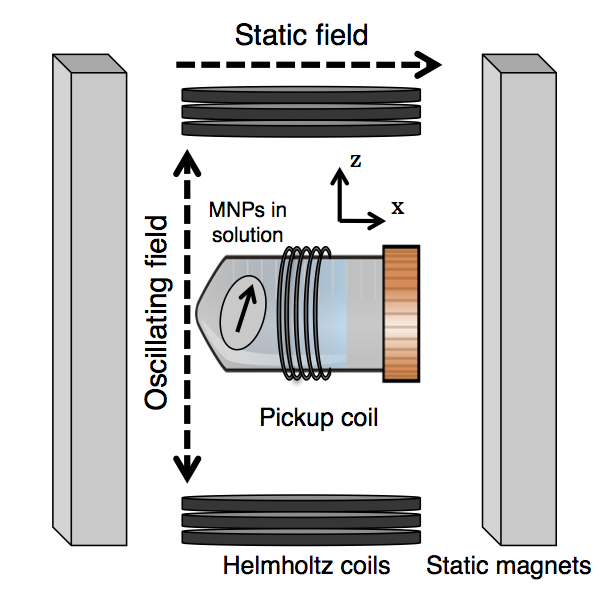}
\caption{A conceptual schematic for static field magnetic nanoparticle spectroscopy (sMPS). An oscillating field is generated by Helmholtz coils and a perpendicular static field by permanent magnets.  Magnetic nanoparticles will rotate in time with the oscillating field, but are guided through the static field direction creating a non-zero average change in flux through the pickup coil.} \label{schema} \end{center} \end{figure}

One of the largest experimental challenges of MPS is that $\mathcal{S}$ is affected by the excitation field. We call this feed-through signal. The MNP signal is orders of magnitude smaller than the driving field; thus, spectroscopy often requires gradiometers or filtering \cite{FILTER}. To address this challenge, we introduce a new method (U.S. Patent Application Serial No. 61/721,378) termed `static field magnetic nanoparticle spectroscopy' (sMPS) that aligns the pickup coil perpendicularly to the excitation field to geometrically decouple the sensing from the drive field. From elementary magnetism the magnetic flux in a current loop, $\phi=\int \mathrm{d}\mathbf{B} \cdot \mathbf{A}$, is zero if the coil is perpendicular to the field. However, if the pickup coil is perpendicular to the excitation field, the average flux from the nanoparticles will also be zero as they are free to rotate in any direction. By adding a static field that is aligned with the pickup coil, the nanoparticles switch orientation in the oscillating field, but are guided through the perpendicular direction, breaking the symmetry of the rotation and ensuring a non-zero magnetization flux in the pickup. A possible apparatus uses Helmholtz coils to drive an oscillating field through the sample, as well as permanent magnets that are large enough to create a uniform static field throughout the sample volume. A small solenoidal pickup coil surrounds the MNP sample and detects MNP magnetization changes perpendicular to the alternating field. A conceptual realization is shown in Fig.~\ref{schema}. In principle, the decoupling provides excellent sensitivity because the signal from the drive field is zero and the output can be amplified to an arbitrarily high level. Additionally, minimizing feed-through signal avoids challenges created by flux canceling with gradiometers or filtering---field inhomogeneity or frequency dependent filtering respectively.

\section{Stochastic simulations of sMPS}

\begin{figure*}[t!] \begin{center} \includegraphics[width=7in]{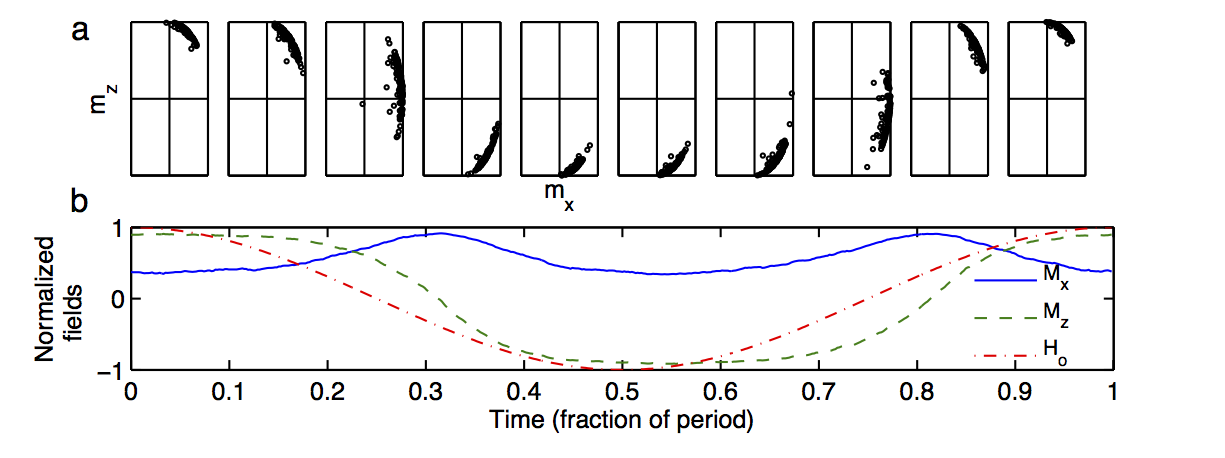} \caption{a) Simulations of one hundred normalized MNP magnetizations projected onto the x-z plane with each dot representing the head of a single nanoparticle magnetization vector. Each of the ten axes is a single time point (arranged in tenths of a period of the oscillating field). The magnetizations switch in the direction of the oscillating field while preferentially rotating through the direction of the static field. b) The average magnetizations $M_x$ and $M_z$ plotted relative to the reference applied oscillating field $H_o$. $M_z$ oscillations follow the reference field, but lag slightly and saturate. In the perpendicular frame, $M_x$ now has twice the frequency of oscillation and also carries information about saturation and phase lag.} \label{mags} \end{center}\end{figure*}

Using previously developed simulations detailed in Reeves and Weaver, we examine the dynamic magnetization of the particles in three-dimensions\cite{DBRSIM}. The model solves Eq.~\ref{sde}, a stochastic Langevin equation for the balance of torques on isotropic, non-interacting Brownian nanoparticles in an overdamped regime \cite{GARPAL}:
\begin{equation} \frac{d\mathbf{m}}{d t}= \frac{\left( \mathbf{m} \times \mathbf{\alpha}\right)  \times \mathbf{m}}{2\tau}  +  \frac{\mathcal{N}\times \mathbf{m}}{\sqrt{\tau}}, \label{sde}\end{equation}
 with the normalized magnetization vector ${\bf m}$,  $\tau$ the Brownian relaxation time as in Eq.~\ref{tau}, and $\alpha$ a unitless field, the ratio of magnetic to thermal energy
\begin{equation}\alpha = \frac{\mu \mathbf{H}}{k_\mathrm{B}T}.\label{alpha}\end{equation}  Note the magnetic field determines the direction of the unitless field and that the magnetic moment magnitude $\mu=M_sV_{\mathrm{co}}$ is a product of saturation magnetization and the iron core volume. The white noise torque $\mathcal{N}$ is implemented as a Wiener process \cite{CHANDRA} and is delta-autocorrelated in time and cartesian directions $i,j\in \hat{x},\hat{y},\hat{z}$ such that
 \begin{equation} \langle\mathcal{N}(t)\rangle=0, \hspace{2mm} \langle\mathcal{N}_i(t)\mathcal{N}_j(t')\rangle=\delta_{ij}\delta(t-t'). \end{equation} 
 To simulate the fields in sMPS, we define
 \begin{equation}  \mathbf{H} = H_s\hat{x} + H_o\cos{w_o t} \hspace{1mm} \hat{z}, \label{sMPSfield}\end{equation} 
where the oscillating field and static field strengths are $H_o$ and $H_s$ respectively. An Euler-Marayuma solver was used \cite{GARD}, and solutions were found from averages of $N=10^4$--$10^6$ simulated particles---i.e., $M=\langle {\bf m} \rangle = \frac{1}{N}\sum_i^N {\bf m}_i $. Particles of diameter 100nm with mass magnetization 70emu/g and 25nm diameter core size were simulated in 100Hz--10kHz magnetic fields on the order of 10mT/$\mu_0$ in conditions near room temperature and water viscosity. 

\begin{figure}[h!]\begin{centering}\includegraphics[width=3.25in]{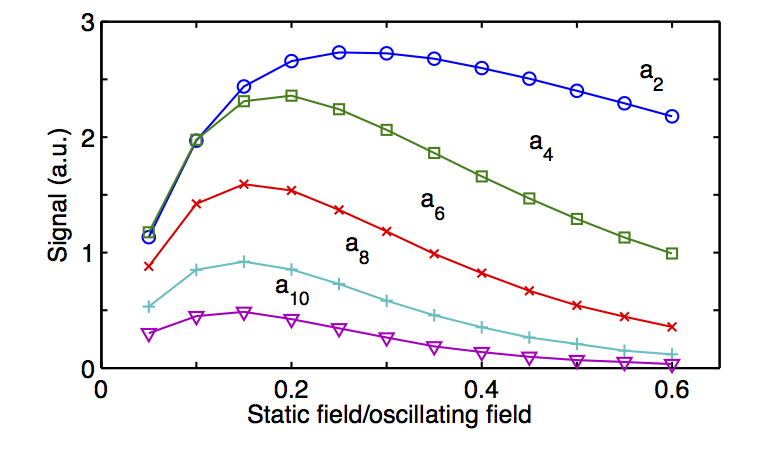} \caption{Simulated harmonics are shown in arbitrary units (a.u.) as a function of static to oscillating field amplitude ratio $\mathcal{H}$ demonstrating the existence of optimizable maximum harmonic signal at $\mathcal{H}_{max}$.} \label{opt}  \end{centering}\end{figure}

Before averaging, we can examine the trajectories of MNP rotations separately. This is visualized with each magnetization as a separate point projected onto the x-z plane. Time slices are presented in Fig.~\ref{mags}a and show that the magnetizations rotate in time with the oscillating field but turn preferentially through the static direction---necessarily creating a non-zero flux in a coil aligned with the static direction. The signal from the MNPs is thus decoupled from the driving field and accessible for sensitive experimental measurement. Because the problem is stochastic, it is more meaningful to discuss statistical moments of the MNPs. Averaged magnetizations in the parallel and perpendicular directions are compared to the reference oscillating field in Fig.~\ref{mags}b. The average magnetization in the perpendicular direction, $M_x$, diminishes as the oscillating field pulls the magnets toward the poles of the z-axis, and $M_x$ peaks slightly after the sinusoidal driving field passes through zero, demonstrating a phase-lag due to the viscous drag. 

As we studied simulations, we found the harmonics depend on the ratio of field magnitudes ($\mathcal{H}=H_s/H_o$). The harmonics are not affected identically by changing $\mathcal{H}$ but as shown in Fig.~\ref{opt}, there is a certain ratio (termed $\mathcal{H}_{max}$) for each harmonic, where the harmonic peaks. Once past the threshold, the static field prevents the MNPs from reaching the poles of the z-axis and the harmonics decrease. Thus the signal strength of the harmonics can be optimized by choosing the correct field ratio. 

\subsection{Relaxation time measurements with frequency scaling}\label{TauW}

Previous experimental work has shown that harmonic amplitudes depend directly on a product of the frequency and relaxation time \cite{JBWK}. Furthermore, this is shown explicitly in the paper of Martens \emph{et al.} through the axially symmetric Fokker-Planck equation \cite{MARTENS}. The full Fokker-Planck equation can be derived from a drift-diffusion assumption\cite{WFB} and written for the distribution of magnetizations $f$ for which the average magnetization $M$ is the first moment:
\begin{equation} 2\tau \frac{\partial f}{\partial t} = \hat{R}\left(\hat{R}-\mathbf{m}\times \alpha \right)f. \label{FP} \end{equation}
where $\hat{R}=\mathbf{m}\times \frac{\partial}{\partial \mathbf{m}}$ is the rotation operator\cite{FELDY}.
Inserting the field of sMPS and employing the change of variables to a unitless time $t'=t/\tau$ leads to
\begin{equation} 2 \frac{\partial f}{\partial t'} = \hat{R}\left(\hat{R}-\mathbf{m}\times  \left(\alpha_s\hat{x}+\alpha_o\hat{z}\cos{\omega_o \tau \hspace{1mm} t'}\right)   \right)f. \label{FP} \end{equation}
 so that the distribution of magnetizations due to an oscillating driving field only a function of the product of $\omega_o \tau$ and the unitless field amplitudes $\alpha_s=\mu H_s/k_BT$ and corresponding $\alpha_o$. The magnetization also will be a function of the same product $M(\omega \tau)$. 

\begin{figure}[h!] \begin{center} \includegraphics[width=3.25in]{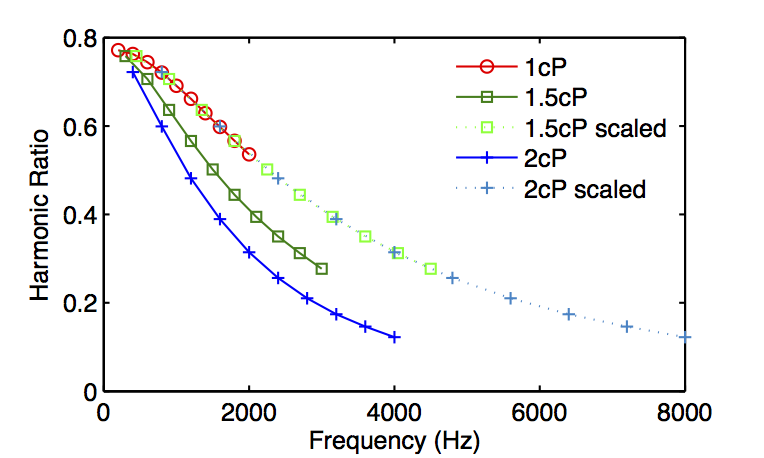} \caption{A scaling of frequency and relaxation time can be used to infer $\tau$ and thus viscosity. A ratio of the fourth over the second harmonic is used, generated from the perpendicular magnetization. Data at increased $\tau$ (simulation with 1.5 and 2 times the original viscosity)  shown before and after scaling by the corresponding amount e.g. $\zeta_f=1.5$ for squares and $\zeta_f=2$ for crosses. Post scaling, the data align onto a single curve.} \label{r42v} \end{center}  \end{figure}

The implication is that two sets of magnetization harmonics with different relaxation times plotted as a function of frequency will be two distinct curves. But, if the frequency axis is scaled by the respective relaxation times, the data map back onto the same curve. This scaling factor $\zeta_f$ can be found from two data sets by least-squares regression, and thus the relaxation time difference from the two sets can be inferred. Allowing the relaxation time to be perfectly Brownian, we can detect changes in viscosity (see Eq.~\ref{tau}). As shown in Fig.~\ref{r42v}, the scaling remains viable when the perpendicular harmonics are considered, corroborating the variable dependence of the full Fokker-Planck equation. In this case, a simulated ratio of fourth to second harmonics is considered $r_{4/2}$. The harmonic signal $\mathcal{S}$ is found by taking the absolute value of the Fourier transform of the magnetization data and multiplying by the harmonic number as in Eq.~\ref{sig}.

\subsection{Temperature measurements by varying the static field amplitude}\label{BsT}

The static field magnitude has a noticeable effect on the nanoparticle magnetizations (see Fig.~\ref{opt}) suggesting that the static field strength could be used as a new variable to examine the properties of the nanoparticles. Measurements decoupled from the oscillating field are important, for instance, because hyperthermia drive coils are made for fixed frequencies and very high amplitudes to have optimal therapeutic consequence \cite{HYPERTHERM}. By measuring MNP spectra at various static field amplitudes, we can determine the reference static field vs. harmonic spectra. Subsequent spectra change with temperature and thus calibrating with the initial reference determines heating.  

The temperature measurement relies on a similar argument to that used in Weaver, Rauwerdink, and Hansen; that the magnetizations and thus the magnetization harmonics $\mathcal{S}$ depend on a ratio of the applied field and the temperature\cite{TEMPEST}. 

We know that if an ensemble of magnetic dipoles is placed in a magnetic field and allowed to evolve for times much longer than a single relaxation time, the average equilibrium magnetization (from Boltzmann statistics) is governed by the Langevin function, $\mathcal{L}(x) = \coth{x}-1/x$. The magnetization of dipoles in a three-dimensional magnetic field is
\begin{equation} \mathbf{M}_{eq} = M_0\mathcal{L}\left(|\alpha|\right)\frac{\mathbf{H}}{|H|}, \end{equation}
where $\mathbf{H}$ is the applied field including direction and $H$ is the magnitude of this field, $M_0=\mu N$ is the total magnetization strength depending on the moment $\mu$ and the number of dipoles $N$. The argument in the Langevin function is the magnitude of the unitless field $\alpha$ defined in Eq.~\ref{alpha}. 

To maintain the scaling relationship between field strength and temperature with a time dependent field $\mathbf{H} \rightarrow \mathbf{H}(t)$, we assume that equilibrium is always maintained. If the frequency of the excitation field is far less than the inverse relaxation time, an adiabatic approximation is justified by the fluctuation dissipation theorem \cite{KUBO}. This is reasonable for Brownian magnetic nanoparticles with relaxation times often on the order of 100$\mu$s---so that even kHz frequencies have periods tens of relaxation times long.

\begin{figure}[ht!] \begin{center} \includegraphics[width=3.25in]{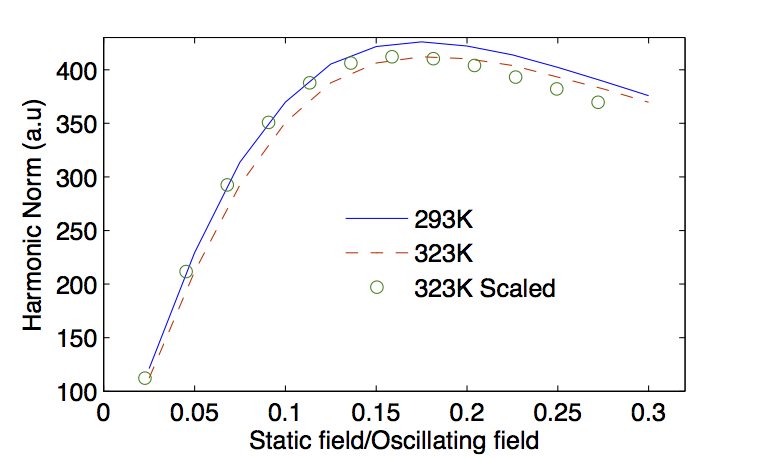}\caption{Simulated harmonic norms for various $\mathcal{H}$ ratios. The same data sets are shown pre- and post-scaling by the corresponding ratio of change in temperature $T_f/T_i$; e.g., the new domain for the second curve is the original static field data points multiplied by $293/323$. Thus, the data below the threshold field ratio align onto the same curve. When the ratio of temperatures is not known in advance, a least-squares fit can be used to find the correct scaling factor, determining the final temperature relative to an initial measurement.} \label{scaleBsT}
\end{center}  \end{figure}

The average magnetization in the perpendicular direction during sMPS can be written in its equilibrium form:
\begin{equation} M_x = M_0\mathcal{L}\left(\frac{\mu H(t)}{k_\mathrm{B}T}\right)\frac{H_s}{H(t)}, \end{equation}
where from Eq.~\ref{sMPSfield}
\begin{equation} H(t)=\sqrt{H_s^2+(H_o\cos{\omega_o t})^2}. \end{equation} 
In the regime where $H_o\gg H_s$ or equivalently, $\mathcal{H}\ll 1$, we make a second approximation by considering the static field to create only a small perturbation on the particle oscillations. Hence, the perpendicular field is only affecting dynamics as the oscillating field nears zero. At this point we have $H(t)\rightarrow H_s$, and thus,
\begin{equation} M_{x} = M_0\mathcal{L}\left(\frac{\mu H_s}{k_\mathrm{B}T}\right), \end{equation}
where the magnetization is clearly a function of the ratio of the static field strength to temperature. Thus, a temperature measurement can be accomplished by varying the static field amplitude and developing a calibration curve of the harmonics. The data at different temperatures appear different until each domain is scaled accordingly by the respective temperatures. Given an initial data set, a least-squares fit can be used to find the correct scaling factor and thus the new temperature of a second data set. Simulations verifying the technique are shown in Fig.~\ref{scaleBsT}.

A few issues do arise. The first is that above the threshold field ratio $\mathcal{H}_{max}$, the scaling breaks down because the perturbation approximation is no longer valid. This is visible in the simulation Fig.~\ref{scaleBsT} as well as the experimental data presented later in Fig.~\ref{expT}. Here the harmonic norm decreases with increasing $\mathcal{H}$ so the temperature scaling fails. Another issue derives from the fact that with typical conditions and MNPs, static fields of the order 1mT/$\mu_0$ lead to unitless field $\alpha$ less than unity. Thus the Langevin function can be approximated as the first few terms of a Taylor series,
\begin{equation} \mathcal{L}(\alpha) \approx \frac{\alpha}{3}-\frac{\alpha^3}{45}+\ldots, \end{equation}
and the magnetization is almost linear in $\alpha$. Thus, the harmonic ratio is no longer a good metric to view subtle changes in magnetization. Instead we can use the harmonic norm \begin{equation} a_{\mathrm{norm}}=\sqrt{\sum_{i}a_i^2}. \end{equation} For the experimental data, it suffices to only use the first few even harmonics which dominate the signal. It may be possible to exploit the higher (e.g., $a_6, a_8$) harmonics to find a usable harmonic ratio. We will explore this possibility in future work. In the case where N\'{e}el relaxation becomes the dominant mechanism, the information derived from the harmonics should still be valid if the relaxation time is much shorter than the period of excitation. Particularly, in hyperthermia applications, the fields can have frequencies of hundreds of kiloHertz so the nanoparticles must have relaxation times shorter than microseconds. This seems to often be the case for instance as demonstrated experimentally and theoretically by Fannin and Charles \cite{FANNIN}. 

\section{Experimental verification of simulated methods \emph{in vitro}}

\subsection{sMPS apparatus}

\begin{figure}[h!] \centering \includegraphics[width=3.25in]{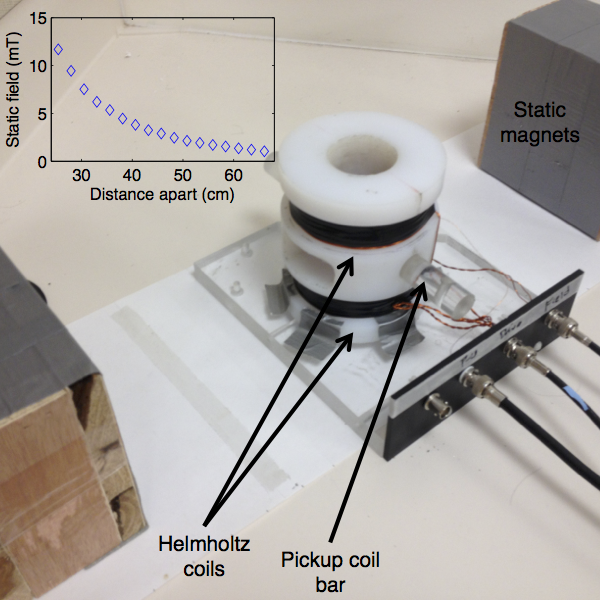}
\caption{A working physical realization of the sMPS apparatus. The static disc magnets are secured in wooden boxes and separated corresponding to the desired static field strength. The Helmholtz pair is wound on a plastic spool and the rod through the middle holds the pickup coil so that fine-tuning the perpendicular angle minimizes feed-through signal. The MNP sample is inserted through the side and fits tightly into the pickup. The static field strength due to the distance between the static magnets is inset.} \label{apps} \end{figure}

A prototype (Fig.~\ref{apps}) modeled after the schematic (Fig.~\ref{schema}) was built to validate the simulated methods \emph{in vitro}. The oscillating field was generated by resonant 3cm Helmholtz coils wound with three-hundred turns of awg 20 copper wire.  The voltage was supplied by a phase-locked amplifier (Zurich HF2LI) amplified by a 1000W audio amplifier (Behringer NU1000) and capacitors were selected with a USB switchboard (Measurement Computing ERB-24) for multiple resonant frequencies. Two niobium disc magnets generated a uniform static field up to 7mT/$\mu_0$ inside the sample region. The graphic inset of Fig.~\ref{apps} shows the static field magnitude per magnet separation distance as measured by a Tesla meter (FW Bell 5100). A 5mm diameter pickup coil of 20 windings of awg 30 magnet wire was mounted on a rotating rod that is used to fine-tune the pickup coil angle, minimizing feed-through signal. From the pickup, a low-noise preamplifier (Stanford Research SR560) band-passed frequencies between 100Hz--10kHz with gain of $10^3$. The amplified signal was directed back into the lock-in amplifier for harmonic analysis with filters set to 48dB/oct and 1Hz bandwidth. The second and fourth ($a_2, a_4$) signal magnitudes are read out simultaneously and each is averaged over 9 seconds. An additional coil monitored the strength of the driving field at the driving frequency with the same parameters except for a shorter 2 second readout. Data collection and excitation field feed-back loop adjustment routines were computer automated with Matlab.

\subsection{Experimental results}

\begin{figure}[h!] \centering \includegraphics[width=3.25in]{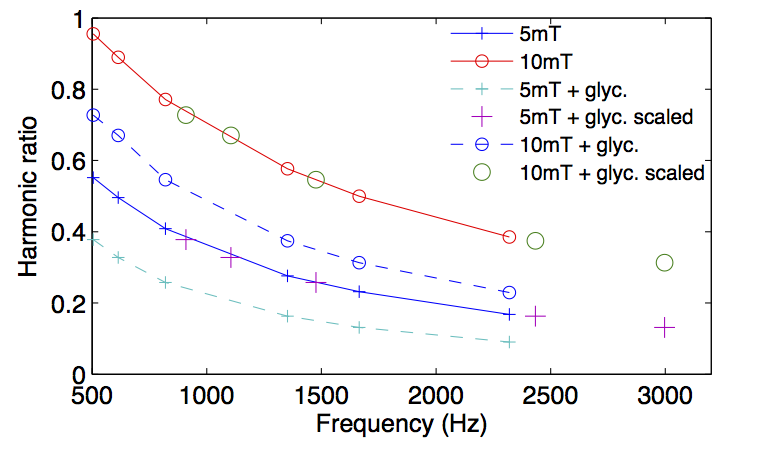}
\caption{Spectra from a MNP sample measured before and after the addition of glycerol. Multiplying the domain of the more viscous sample data by the ratio of viscosities scales the data back onto the original curve. With a unknown amount of glycerol added, it is possible to measure the a change in viscosity by least-squares regression to find the domain scaling factor $\zeta_f$.}
\label{exptau} \end{figure}

Experiments with the new apparatus were designed to verify both scaling relationships: frequency with relaxation time $M(\omega\tau)$; and static field with temperature $M(H_s/T)$. In both cases, samples with varying concentrations of MNPs (Micromod 100nm diameter BNF starch iron oxide) were prepared in buffer in 200$\mu$L test tubes. 

For the relaxation time measurement shown in Fig.~\ref{exptau}, a series of resonant frequencies between 400Hz--2400Hz were used and the static field was held at 1mT/$\mu_0$. The oscillating field was constantly monitored and updated to stay at the desired strength. A sample containing 150$\mu$g of iron in 100$\mu$L of PBS was measured at two oscillating field amplitudes. Glycerol was added, amounting to 22\% by weight. At room temperature, this changes viscosity by a factor of 1.81 \cite{GLYC}. The data analysis method is detailed in Sec.~\ref{TauW}. By taking a harmonic ratio at each frequency, a characteristic curve can be developed. When the viscosity is increased, the characteristic curve drops because the thicker liquid impedes rotation, decreasing MNP saturation and therefore diminishing the higher harmonics. Scaling the domain by the correct factor (the change in the relaxation time $\zeta_f$) aligns the data back onto the original curve. A least-squares regression determined the scaling factor to be $\zeta_f=1.8\pm0.2$, so that the predicted viscosity change by weight is accurate to within a standard deviation.

\begin{figure}[h!] \centering \includegraphics[width=3.25in]{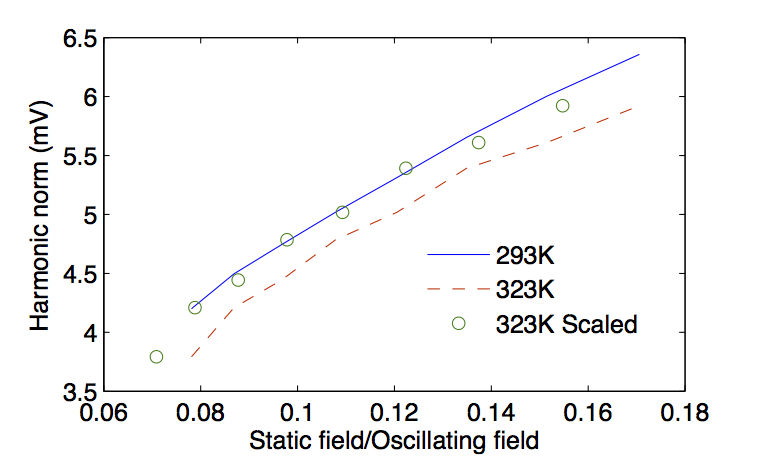}
\caption{MNP harmonics depend on a ratio of $H_s/T$. Originally, two different temperature samples have different harmonic spectra. By dividing the domain of each data set by the respective temperature factor, a single curve results. This demonstrates the important result that temperature measurements of MNPs are possible by only varying the static field.} \label{expT} \end{figure}

The temperature measurements shown in Fig.~\ref{expT} are accomplished by changing the magnitude of the static field, as explained in Sec.~\ref{BsT}. For this experiment, an identical MNP sample was made. The excitation field was 10mT/$\mu_0$ at 490Hz and varying static field amplitudes were used from 0.5 to 2mT/$\mu_0$. We then heated the same sample to 323K in a dry bath and remeasured the harmonic spectra with the same magnetic fields.

\subsection{Comments on the sensitivity of the current apparatus}

We have seen that the current prototype is capable of repeating spectroscopic experiments with the signal now the even harmonics of the magnetization perpendicular to the oscillating field. The geometric decoupling from the drive coil minimizes feed-through signal from the excitation field. Thus the MNP signal can be amplified as needed so that signals from MNPs surpass the lock-in amplifiers' baseline detection limits. In the experiments, the temperature and viscosity measurement methods were demonstrated with iron concentrations much above the minimum sensitivity and thus the error bars are negligible. The current noise sources are mechanical vibration of the drive coil coupling to the pickup, and drive field inhomogeneities. We expect that with small improvements (e.g., a cooling system for the excitation coil to avoid field fluctuations, larger drive coils compared to the pickup coil geometry, and more precise alignment of static magnets to avoid field gradients) the sensitivity could eventually only be limited by the small noise generated by the pre-amplifier.


\section{Conclusions}

The concept of static field magnetic nanoparticle spectroscopy (sMPS) is introduced as a way to measure magnetic nanoparticle (MNP) properties. By turning the pickup coil perpendicular to the excitation field and adding a static field aligned to the pickup, it is possible to detect magnetization flux from the MNPs, geometrically decoupling the sensing coil from the excitation fields while still determining nanoparticle dynamics. The decoupling means that experimental challenges due to large feed-through voltages from the excitation field can be nearly eliminated and complicated signal processing procedures like filtering or gradiometers can be avoided. With minimized background signal, the MNP flux voltages can be amplified arbitrarily to achieve high sensitivity. This advances the resolution and minimum molecular concentrations possible in the widespread technology of biological sensing using MNP probes.

Stochastic simulations effectively described three-dimensional rotations of magnetic nanoparticles in various field geometries. The simulations show that the harmonic spectra of MNPs in the perpendicular direction allow measurements including relaxation times from frequency scaling as well as temperatures from a varying static field. The new method to measure MNP temperature is developed and shows promise for concurrent temperature measurements during MNP hyperthermia (a promising cancer treatment for which effective therapy and patient safety relies on accurate real-time monitoring of the temperature of magnetic particles). The perpendicular measurements are interesting because previous techniques face challenges from interference between the sensing and the necessarily rigid, high-power hyperthermia excitation field. 

It should be noted that the frequency regimes studied here are not the same as those employed in hyperthermia. Furthermore, although the heating mechanism of hyperthermia is not fully understood, it is expected to be caused by a combination of Brownian and N\'{e}el relaxation. In either mechanism however, if the period of the oscillating field is longer than the relaxation time of the particle, the equilibrium assumptions of Sec.~\ref{BsT} should remain valid, and thus the temperature measurements are theoretically possible. 

An apparatus was designed and built that validated the simulations and especially showed that the static field temperature measurements are possible, increasing the motivation to attempt a multi-modality apparatus combining sMPS and hyperthermia.

\begin{acknowledgments} This work was supported by an NIH-NCI grant 1U54CA151662-01 and the Neukom Fellowship of Dartmouth College.\end{acknowledgments}


\begin{thebibliography}{99}
	
\bibliographystyle{unsrt}

\bibitem{KOH} Koh I \& Josephson L, Magnetic nanoparticle sensors. \emph{Sensors}, {\bf9(10)}:8130 (2009).

\bibitem{HAUN} Haun J B, \emph{et al.}  Magnetic nanoparticle biosensors. \emph{WIRES: Nano}, {\bf4972(3)}:291 (2010).

\bibitem{SEIKO} Weaver J B, \emph{et al.} Magnetic nanoparticle quantitation with low frequency magnetic fields: compensating for relaxation effects. \emph{Nanotechnology}, {\bf 24}:325502 (2013).

\bibitem{XJ} Zhang X, \emph{et al.} Molecular sensing with magnetic nanoparticles using magnetic spectroscopy of nanoparticle Brownian motion. \emph{Biosensors and Bioelectronics}, {\bf 50} (2013).

\bibitem{PANK} Pankhurst Q A, Connolly J, Jones S K, \& Dobson J, Applications  of  magnetic  nanoparticles  in  biomedicine. \emph{J. Phys. D: Appl. Phys.}, {\bf50036(13)}:R167 (2003).

\bibitem{TEMPEST} Weaver J B, Rauwerdink A M, \& Hansen E, Magnetic nanoparticle temperature estimation. \emph{Med. Phys.}, {\bf 36(5)}:1822 (2009).

\bibitem{VISC} Rauwerdink A M \& Weaver J B, Viscous effects on nanoparticle magnetization harmonics. \emph{J. Magn. Magn. Mater.}, {\bf322}:609-614 (2010).

\bibitem{RIGID} Weaver J B, \emph{et al.} Micro-rheology: evaluating the rigidity of the microenvironment surrounding antibody binding sites. \emph{Proc. SPIE}, 76261J (2010).

\bibitem{HERGT} Hergt R, Dutz S, Muller R, \& Zeisberger M, Magnetic particle hyperthermia: nanoparticle magnetism and materials development for cancer therapy. \emph{J. Phys.: Condens. Matter}, {\bf18(38)}:S2919 (2006).

\bibitem{JORDAN} Jordan A, Scholz R, Wust P, Fahling H, \& Felix R, Magnetic fluid hyperthermia (MFH): Cancer treatment with AC magnetic field induced excitation of biocompatible superparamagnetic nanoparticles. \emph{J. Magn. Magn. Mater.}, {\bf 201(13)}:413 (1999).

\bibitem{HYPERTHERM} Giustini A J, \emph{et. al.} Magnetic nanoparticle hyperthermia in cancer treatment. \emph{Nano LIFE}, {\bf 17(01)} (2010).

\bibitem{BIHAN} Bihan L, Delannoy J, \& Levin R L, Temperature mapping with MR imaging of molecular diffusion: application to hyperthermia. \emph{Radiology}, {\bf 171(3)}:853-857 (1989).

\bibitem{TCC} Cetas T C, Will thermometric tomography become practical for hyperthermia treatment monitoring? \emph{Cancer Res.}, {\bf44}:4805-4808 (1984).

\bibitem{MONITOR} Weaver J B, The use of magnetic nanoparticles in thermal therapy monitoring and screening: Localization and imaging. \emph{J. Appl. Phys.}, {\bf111(7)}:07B317 (2012).

\bibitem{WFB} Brown W F, Thermal fluctuations of a single domain particle. \emph{Phys. Rev.}, {\bf 130(5)} (1963).

\bibitem{EIN} A Einstein. Investigations on the theory of the Brownian movement. Dover, Mineola, MN (1956).

\bibitem{FILTER} Graeser M, Knopp T, Gruttner M, Sattel T F, \& Buzug T M, Analog receive signal processing for magnetic particle imaging. \emph{Med. Phys.}, {\bf40(4)}:042303 (2013).

\bibitem{DBRSIM} Reeves D B \& Weaver J B, Simulations of magnetic nanoparticle brownian motion. \emph{J. Appl. Phys.}, {\bf 112}:124311, (2012).

\bibitem{GARPAL} Garcia-Palacios J \& Lazaro F, Langevin-dynamics study of the dynamical properties of small magnetic particles. \emph{Phys. Rev. B}, {\bf58}:14937-14958 (1998).

\bibitem{CHANDRA} Chandrasekhar S, Stochastic problems in physics and astronomy. \emph{Rev. Mod. Phys.}, {\bf15(1)} (1943).

\bibitem{GARD} T C Gard. Introduction to Stochastic Differential Equations. Springer, Berlin (2003).

\bibitem{JBWK} Weaver J B \& Kuehlert E, Measurement of magnetic nanoparticle relaxation time. \emph{Med. Phys.}, {\bf 39(5)}:1-6 (2012).

\bibitem{MARTENS} Martens M A, \emph{et al.} Modeling the brownian relaxation of nanoparticle ferroßuids: Comparison with experiment. \emph{Med. Phys.}, {\bf40(2)}:022303 (2013).

\bibitem{KUBO} Kubo R, The fluctuation dissipation theorem. \emph{Rep. Prog. Phys.}, {\bf29}:255 (1966).

\bibitem{GLYC} Sheely M L, Glycerol viscosity tables. \emph{Ind. Eng. Chem.}, {\bf24(9)}:1060-1064 (1932).

\bibitem{FANNIN} Fannin P C \& Charles S W, On the calculation of the Neel relaxation time in uniaxial single-domain ferromagnetic particles \emph{J. Phys. D: Appl. Phys.} {\bf 27} 185 (1994).

\bibitem{FELDY} Felderhof B U \& and Jones R B, Nonlinear response of a dipolar system with rotational diffusion to an oscillating field. \emph{J. Phys. Condens. Matter} {\bf15}, S1363 (2003).

\end{thebibliography}
\end{document}